\documentclass[
11pt,%
final,%
notitlepage,%
oneside,%
onecolumn,%
nobibnotes,%
nofootinbib,%
superscriptaddress,%
noshowpacs,%
centertags]%
{revtex4}

\usepackage{graphicx}
\usepackage[cp1251]{inputenc}

\begin{document}
\selectlanguage{english}

\title{Color Variability of the Blazar AO 0235+16}\thanks{Astronomy Reports, 2007, Vol. 51, No. 11, pp. 882–890. \copyright Pleiades Publishing, Ltd., 2007}

\author{\bf\firstname{V.~A.}~\surname{Hagen-Thorn}}
\affiliation{Sobolev Astronomical Institute, St. Petersburg State University}
\author{\bf\firstname{V.~M.}~\surname{Larionov}}
\affiliation{Sobolev Astronomical Institute, St. Petersburg State University}
\affiliation{Main (Pulkovo) Astronomical Observatory, Russian Academy of Sciences}
\author{\bf\firstname{C.~M.}~\surname{Raiteri}}
\affiliation{INAF—Osservatorio Astronomico di Torino}
\author{\bf\firstname{E.~I.}~\surname{Hagen-Thorn}}
\affiliation{Main (Pulkovo) Astronomical Observatory, Russian Academy of Sciences}
\affiliation{Sobolev Astronomical Institute, St. Petersburg State University}
\author{\bf\firstname{A.~V.}~\surname{Shapiro}}
\affiliation{Sobolev Astronomical Institute, St. Petersburg State University}
\author{\bf\firstname{A.~A.}~\surname{Arkharov}}
\affiliation{Main (Pulkovo) Astronomical Observatory, Russian Academy of Sciences}
\author{\bf\firstname{L.~O.}~\surname{Takalo}}
\affiliation{Tuorla Observatory, University of Turku}
\author{\bf\firstname{A.}~\surname{Sillanp\"a\"a}}
\affiliation{Tuorla Observatory, University of Turku}


\begin{abstract}
Multicolor (UBVRIJHK) observations of the blazar AO 0235+16 are analyzed. The light
curves were compiled at the Turin Observatory from literature data and the results of observations obtained
in the framework of the WEBT program (http://www.to.astro/blazars/webt/). The color variability of the
blazar was studied in eight time intervals with a sufficient number of multicolor optical observations;
JHK data are available for only one of these. The spectral energy distribution (SED) of the variable
component remained constant within each interval, but varied strongly from one interval to another. After
correction for dust absorption, the SED can be represented by a power law in all cases, providing evidence
for a synchrotron nature of the variable component. We show that the variability at both optical and IR
wavelengths is associated with the same variable source.
\end{abstract}

\maketitle

\section{Introduction}

AO 0235+16 has recently been one of the most
intensively studied BL Lac objects~\cite{raiteri2001, raiteri2005, raiteri2006,junkkarinen2004}. Raiteri et al.~\cite{raiteri2001, raiteri2005, raiteri2006} present a virtually complete bibliography with
a description of the object’s parameters, and we will
not reproduce them here. We note only that the variability
amplitude of AO 0235+16 reaches $5^{m}$, and its
observed color is anomalously red for blazars.
An absorption system with $\mathrm{z = 0.524}$ is seen in the
spectrum of AO 0235+16 ($\mathrm{z_{em} = 0.94}$). Junkkarinen
et al.~\cite{junkkarinen2004} showed that the region responsible for this
absorption system lines contains a large amount of
dust that leads to the strong reddening. Therefore,
all estimates of spectral indices and other parameters
obtained before ~\cite{junkkarinen2004} should be revised. In particular,
this is true of the data of Hagen-Thorn et al.~\cite{hagenthorn1990}
(although their conclusion concerning the large color
variations of the sources, which are responsible for
variability with a characteristic time scale of about a
year, remains valid).
The color variability of AO~0235+16 is discussed
in~\cite{raiteri2006}. However, we believe that a more detailed analysis
is needed, using the technique described in~\cite{hagenthorn1999},
which has been applied to the color variability of
blazars in many studies (see, for example,~\cite{hagenthorn2006a, hagenthorn2006b})).

\section{OBSERVATIONS AND TECHNIQUE OF THE ANALYSIS}

Here, we use the $UBVRI$ (optical) and $JHK$
(near-IR) light curves compiled by C.M. Raitieri at
the Torino Observatory, based on published data and
data obtained as a result of combining and comparing
observations taken as part of the WEBT Program
(Whole Earth Blazar Telescope; http://www.to.astro/
blazars/webt/), in which the authors take part. Regular
IR observations started in 2002; optical data
are available since 1975, when the radio source was
identified with a star-like object.
Figure 1 presents light curves in one optical ($R$)
and one IR ($K$) band as examples (nightly averaged
fluxes are given, which we will also use in our
subsequent analysis, thereby eliminating very rapid
variability in favor of studying color variations on time
scales exceeding a day). We transformed the magnitudes
into fluxes using the calibration of Mead et al.~\cite{mead1990}.

\begin{figure}[htb]
\includegraphics[width=12cm]{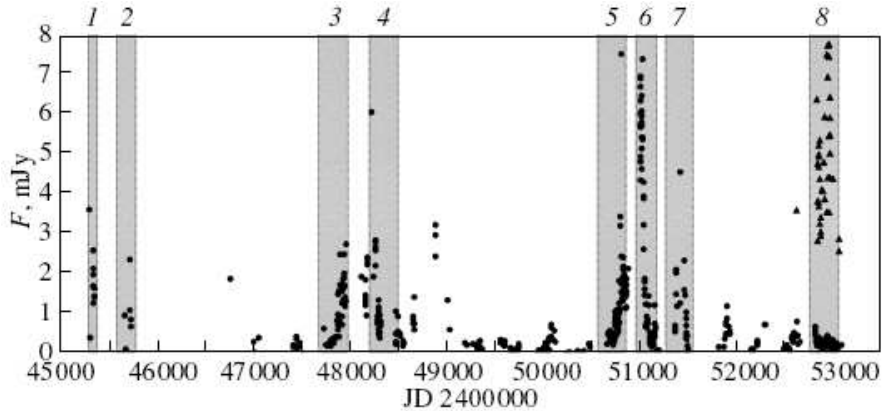}
\caption{\it Light curve of AO 0235+16 in the $R$ (points) and $K$ (triangles) bands. Time intervals within which the color variability
was studied are indicated.}
\label{fig:1}
\end{figure}

The idea of our technique for analyzing color variability
is to construct flux–flux diagrams for pairs
of bands (or, to be more precise, the flux density is
considered). If the color characteristics of the variable
component are unchanged, the points in the diagrams should lie on straight lines, whose slopes yield the
ratios of the fluxes of the {\it variable} component in the
bands considered. In this way, multicolor observations
of variability can be used to derive the relative
spectral energy distribution (SED) of the variable
source.

It is convenient to determine the SED of the variable
source separately for optical and IR wavelengths.
These distributions can then be joined using a flux–
flux diagram comparing fluxes in one of the optical and one of IR bands.

The most noteworthy advantage of the method is
that obtaining the relative SED of a variable component
does not require previous determination of its
contribution to the combined observed radiation.

\section{THE RESULTS OF THE ANALYSIS OF THE COLOR VARIABILITY}

As was shown in our earlier study~\cite{hagenthorn1990}, the color
characteristics of the variable component of
AO~0235+16 can vary, and rather rapidly. Accordingly,
we chose eight time intervals, within which
we determined the relative SED of the variable
component (the light curves indicate that it is the
variable component that determines the photometric
behavior of the object in these intervals, whose edges
are indicated in Fig.~1).
A pairwise comparison of the observed fluxes in
different spectral bands indicates that they are related
linearly (seen from the fact that the correlation coefficients
are close to unity), so that the SED of the
variable component within each time interval may be
considered as constant. We carried out such comparisons
separately for optical and IR wavelengths,
selecting $R$ and $K$ as the primary bands.

\begin{figure}[htb]
\includegraphics[width=6cm]{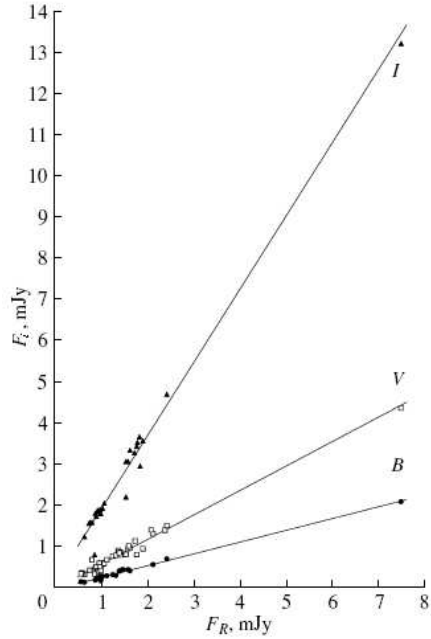}
\caption{\it Flux–flux diagrams for interval 5.}
\label{fig:2}
\end{figure}

\begin{figure}[htb]
\includegraphics[width=6cm]{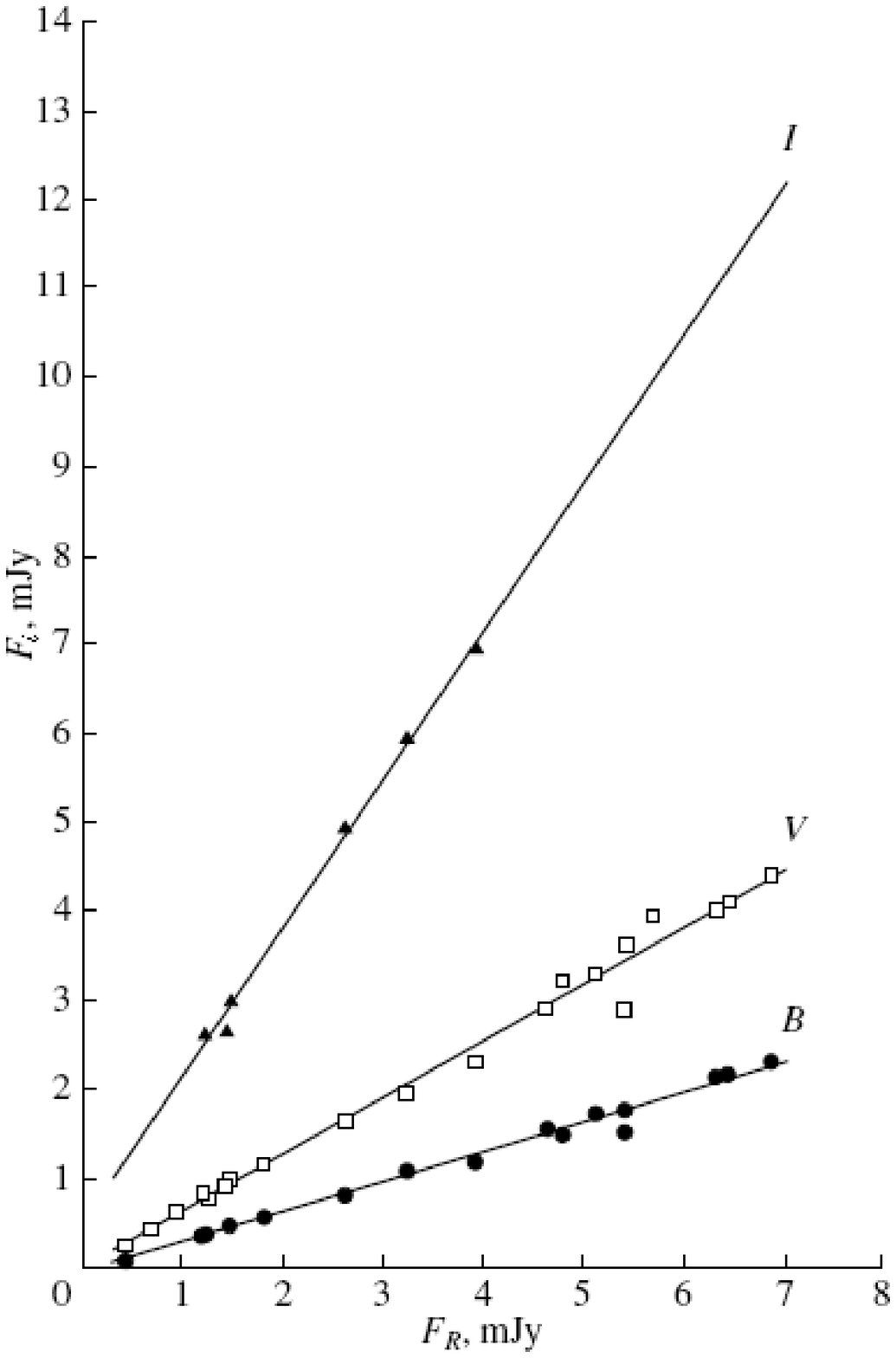}
\caption{\it Flux–flux diagrams for interval 6.}
\label{fig:3}
\end{figure}

\begin{figure}[htb]
\includegraphics[width=6cm]{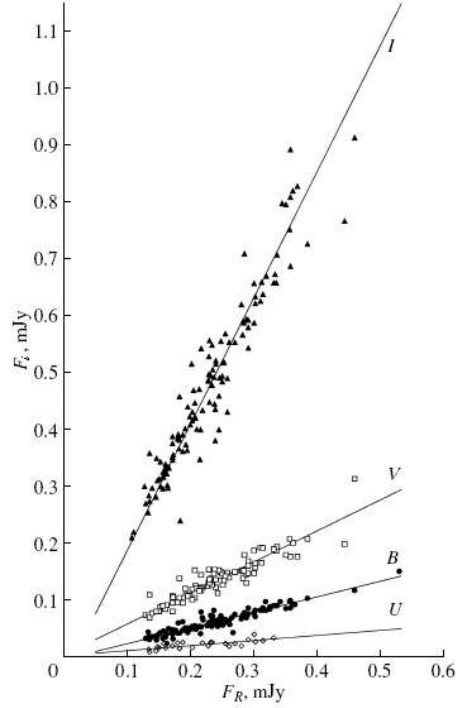}
\caption{\it Flux–flux diagrams for interval 8 at optical wavelengths.}
\label{fig:4}
\end{figure}

\begin{figure}[htb]
\includegraphics[width=6cm]{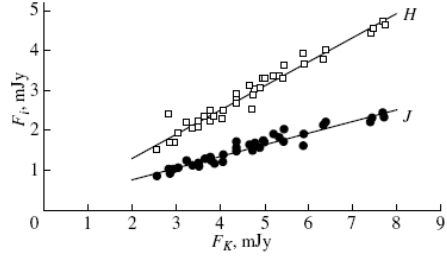}
\caption{\it Flux–flux diagrams for interval 8 at infrared wavelengths.}
\label{fig:5}
\end{figure}

To avoid overburdening the paper, we will present
detailed results for three most interesting intervals,
and only final results for the remaining ones.
Figures~2–5 present the flux–flux diagrams for
intervals 5 (JD 2450600–2450900, 1997–1998),
6 (JD 2451000–2451200, 1998–1999) and 8 (JD 2452800–2453100, 2003). 

\begin{figure}[htb]
\includegraphics[width=6cm]{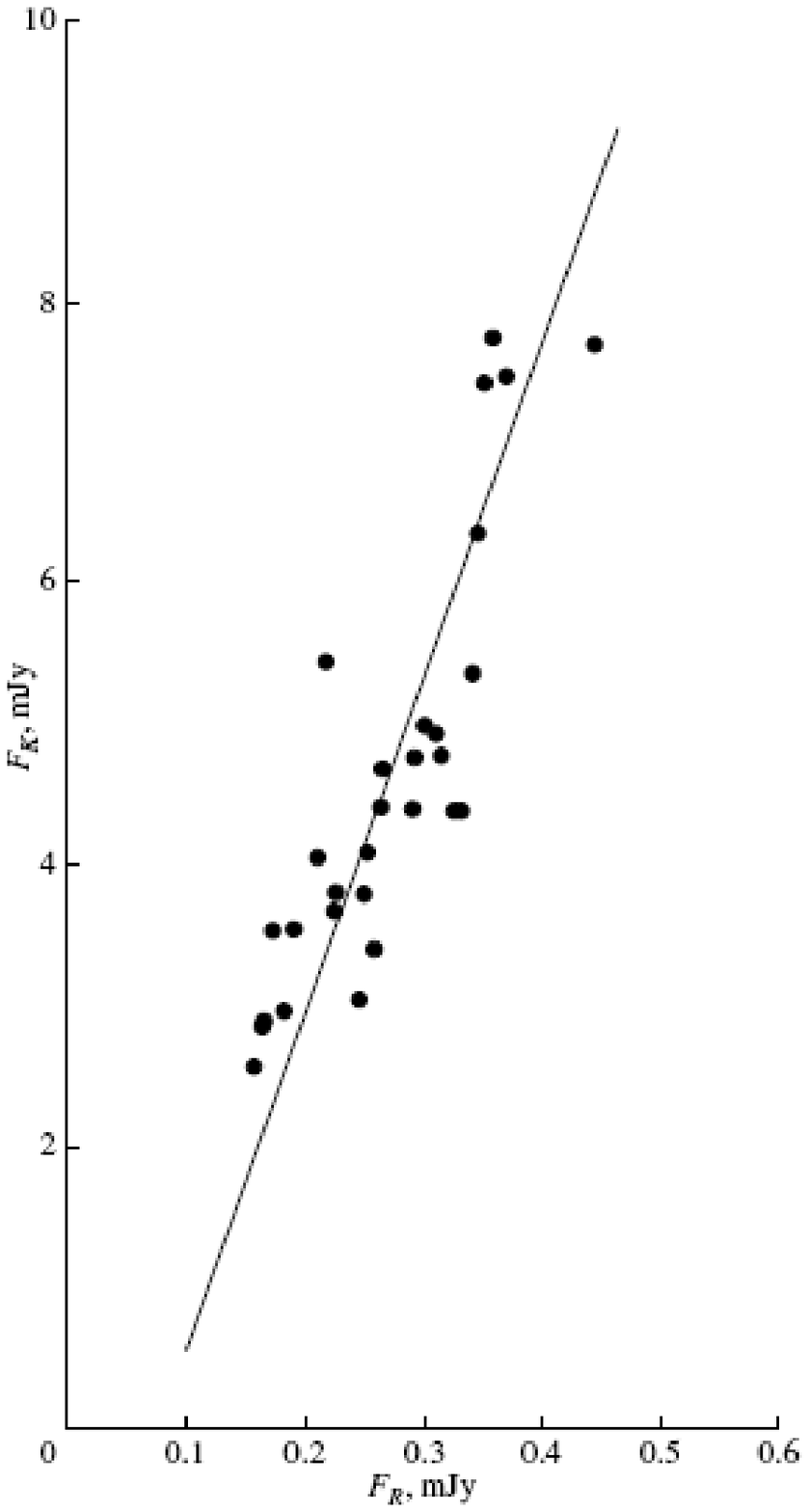}
\caption{\it Flux–flux diagrams for interval 8 and the $R$ and $K$ bands.}
\label{fig:6}
\end{figure}

Figure~6 shows
the flux–flux diagram for R and K bands, used to
construct joint optical and IR spectrum in interval 8
(the only interval for which there are IR data).
The slopes of the lines derived via an orthogonal regression
fit~\cite{wald1940} (this technique should be used
when both compared values display random errors)
are presented in the fifth columns of Tables~1–3,
along with their $1\sigma$ errors. These yield the observed
relative SED of the variable component. (The first
four columns of these tables present (1) band, (2) logarithm
of the corresponding frequency, (3) correlation
coefficient, and (4) number of points in the graphs.)
An important stage in the analysis is correcting
the observed distribution for absorption by intervening
dust clouds. This correction is introduced by multiplying
the values in the fifth columns of Tables~1–3
by the factor $C_{iR}$ ($C_{iK}$ for IR bands), derived from the absorption data taken from the last column of Table~5 in~\cite{raiteri2005}, where all sources of absorption are taken
into account in accordance with~\cite{junkkarinen2004}. The absorption values are presented in the sixth columns of Tables~1–3, and the factors $C_{iR}$ ($C_{iK}$) in the seventh columns.

The $(F_i/F_R)^{corr}_{var}$ ratios, corrected for absorption, which give the relative SED of the variable component, are presented in the eighth columns of Tables~1–2 and the eighth and ninth columns of Table~3. The last columns of Tables~1–3 contain the logarithms of
these ratios, which are plotted against $\log\nu$ in Figs.~7 and 8. Figure~8 separately presents the optical and IR SEDs and the joint SED for the total range from $K$ to $U$.

\begin{figure}[htb]
\includegraphics[width=6cm]{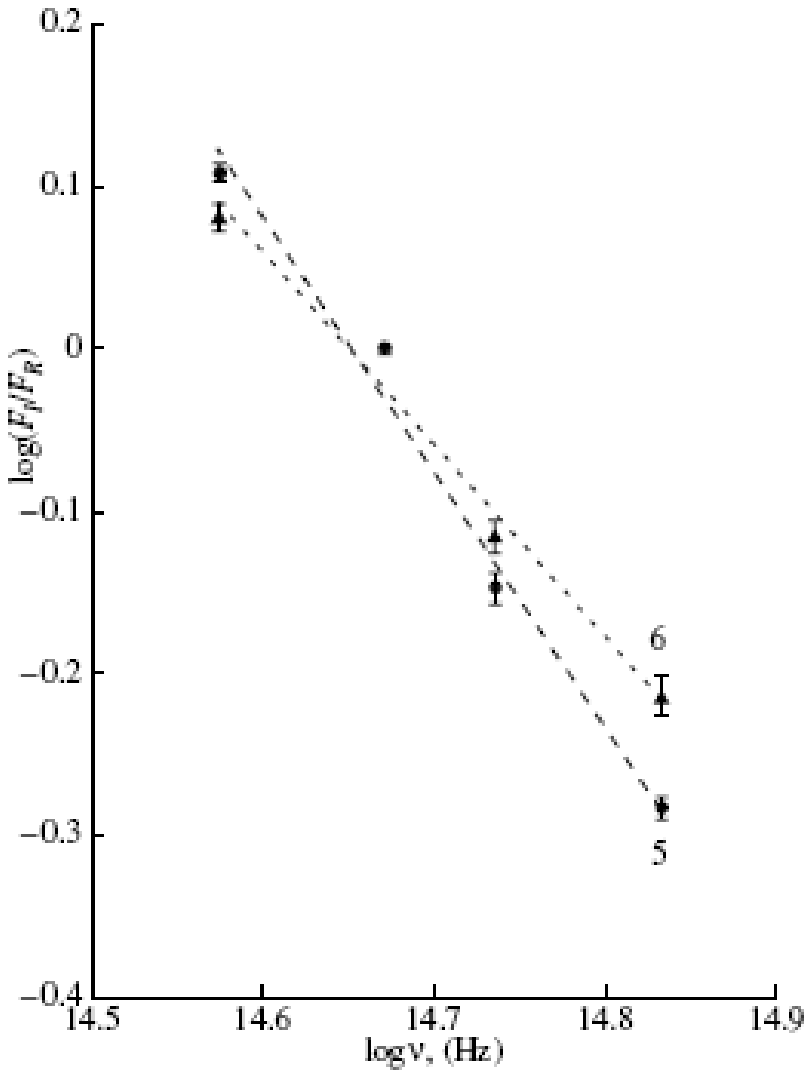}
\caption{\it Spectra of variable sources in intervals 5 and 6.}
\label{fig:7}
\end{figure}

\begin{figure}[htb]
\includegraphics[width=6cm]{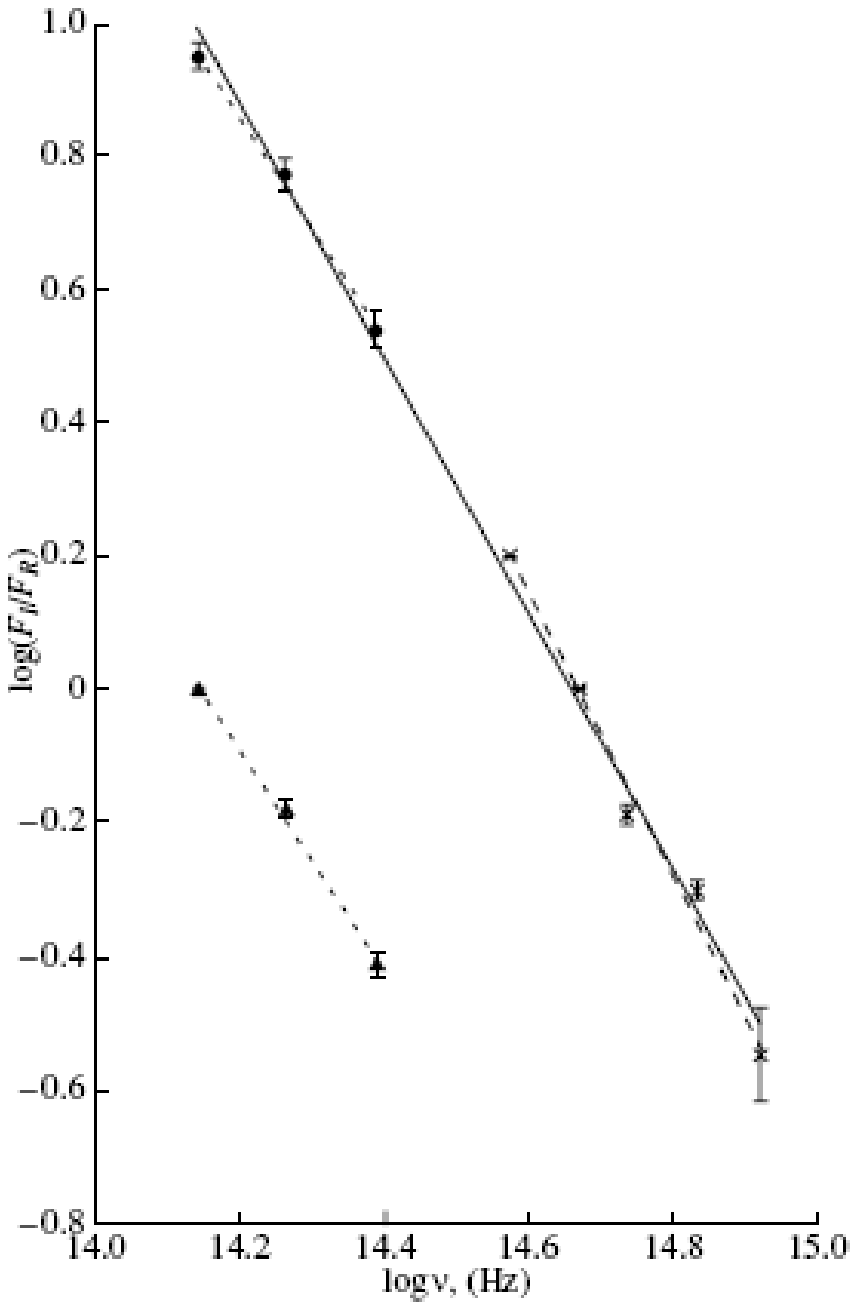}
\caption{\it Spectrum of the variable source in interval 8. The IR (triangles) and optical (crosses) spectra and the ``joint'' spectrum
(points and solid line) are plotted separately.}
\label{fig:8}
\end{figure}

Figures~7 and 8 show that the dots lie along
straight lines fairly well; i.e., the SED of the variable
component follows the power law $F_{\nu} \sim \nu^{\alpha}$, where the
spectral index $\alpha$ is given by the slopes of the lines
in the graphs, which were derived via least-squares fitting. The $\alpha$ values for all eight time intervals are
presented along with their $1\sigma$ errors in the third column
of Table~4 (the second column indicates the
interval, the fourth the correlation coefficient between $log (F_i/F_R)^{corr}_{var}$ and $\log {\nu}$, and the fifth and sixth are the
average and maximum fluxes in the $R$ band). For uniformity,
the $\alpha$ value found from the optical data is presented for interval 8. In this interval, the spectral index $\alpha = -1.66 \pm 0.12$ was found for the IR wavelengths,
with a correlation coefficient of $r=0.99$, while we find
$\alpha = -1.92 \pm 0.05$ with $r=0.99$ for the joint SED

\section{DISCUSSION AND CONCLUSIONS}

The power spectrum and high polarization usually
observed in AO~0235+16 (for example, according to
our unpublished data, it exceeded 30\% in December
2006) provide clear evidence for a synchrotron nature
for the variable sources responsible for this activity.

We can see from Figs.~2–5 that the lifetimes of
the variable sources could be hundreds of days, but
apparently do not exceed a year, since the corresponding
spectral indices for different seasons differ
dramatically (Table~4). This is particularly striking
for intervals 5 and 6. Figure~1 indicates that the flux
increases in interval 5, while it decreases in interval 6. Unfortunately, a seasonal gap in the optical light
curve lies between intervals 5 and 6. One might think
that we are dealing here with a single event (Fig.~8
from~\cite{raiteri2001}, which presents radio light curves, does not seem to contradict this). However, the spectrum is steeper in interval 5, when the flux increases, than in interval 6, when it decreases. Current evolutionary mechanisms for variable sources predict the opposite behavior. Therefore, we conclude that in intervals 5 and 6 different events were observed.

Figure~9 compares the spectral indices with the
average values for a given time interval and maximum
$R$ flux, according to the data in Table~4. In the first case, a random cloud of points is observed;
however, weak ($r=0.2$) correlation is observed for
the maximum flux: harder (flatter) spectra correspond to higher maximum fluxes in an event. Such behavior
(expected from general considerations) was already
noted by Hagen-Thorn et al.~\cite{hagenthorn1990} for the blazar OJ~287.

\begin{figure}[htb]
\includegraphics[width=6cm]{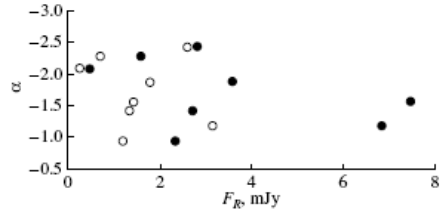}
\caption{\it Comparison of the spectral indices of the variable source at optical wavelengths with the $R$ flux.Open circles correspond
to $\overline F_R$, and filled circles to $F_R^{max}$.}
\label{fig:9}
\end{figure}

Let us now consider Fig.~8. We can clearly see
that there is no discontinuity between the data for
the optical and IR wavelengths in the joint SED.
The optical spectrum is slightly steeper, possibly indicating a high-frequency cutoff. However, the data
for shorter wavelengths (see Fig.~10 in~\cite{raiteri2005}) do not
support this possibility. Such a small break in the
spectrum could be due to a small underestimation
of the absorption (this is quite possible in our case).
Taking this into account, as well as the fact that the
difference between the optical and IR spectral indices
does not exceed the $2\sigma$ uncertainties, we suggest that the variability in both spectral regions was associated with the same variable source, and that the joint SED represents the relative SED in time interval 8 over the entire frequency range from K to U. The spectral
index ($\alpha = -1.92$) that we have derived is rather common for blazars.

In conclusion, we note that, due to the pronounced
differences between the SEDs of the variable component
in different time intervals, a comparison of
the observed color indices and brightnesses for the
entire set of observational data contains little information.
Moreover, such a comparison could lead to
the erroneous conclusion that the SED of the variable
component varies during an outburst.

\begin{acknowledgments}
This work was supported by the Russian Foundation
for Basic Research (project code 05-02-17562). V.A.H.-T. and E.I.H.-T. thank Tuorla Observatory for hospitality. L.O.T. and A.S. are grateful to the Academy of Sciences of Finland, which also supported the study. C.M.R. thanks the Italian Space
Agency, which supported this work in the frame of
Contract ASI-INAF I/023/05/0. Observations with the AZT-24 Telescope are made in accordance with
an agreement between the Main (Pulkovo) Observatory
of the Russian Academy of Sciences and the
Rome and Teramo Observatories (Italy).
\end{acknowledgments}

\newpage

\newpage
\begin{table}
\caption{Results for interval 5 (JD 2450600–2450900, 1997–1998))
}
\begin{center}
\begin{tabular}{ccccccccc}
\hline

Band&$\log \nu$&$r_{iR}$&n&$(F_i/F_R)^{obs}_{var} \pm 1\sigma $&$A_{i},m$&$C_{iR}$&
   $(F_i/F_R)^{corr}_{var} \pm 1\sigma$&$\log (F_i/F_R)^{corr}_{var} \pm 1\sigma$\\
\hline
B & 14.833 & 0.99 & 21 & $0.287 \pm 0.005$ & 1.904 & 1.810 & $0.519 \pm 0.009$&$-0.284 \pm 0.007$\\
V & 14.736 & 0.99 & 35 & $0.596 \pm 0.013$ & 1.473 & 1.194 & $0.711 \pm 0.016$&$-0.148 \pm 0.010$\\
R & 14.670 & -    & -  & 1.000           & 1.260 & 1.000 & 1.000 &          0.000\\
I & 14.574 & 0.99 & 25 & $1.783 \pm 0.022$ & 0.902 & 0.719 & $1.282 \pm 0.016$& $0.108 \pm 0.005$\\

\hline
\end{tabular}
\end{center}
\end{table}

\begin{table}
\caption{Results for interval 6 (JD 2451000–2451200, 1998–1999))
}
\begin{center}
\begin{tabular}{ccccccccc}
\hline

Band&$\log \nu$&$r_{iR}$&n&$(F_i/F_R)^{obs}_{var} \pm 1\sigma $&$A_{i},m$&$C_{iR}$&
   $(F_i/F_R)^{corr}_{var} \pm 1\sigma$&$\log (F_i/F_R)^{corr}_{var} \pm 1\sigma$\\
\hline
B & 14.833 & 0.99 & 16 & $0.337 \pm 0.009 $& 1.904 & 1.810 & $0.610 \pm 0.016$&$-0.215 \pm 0.012$\\
V & 14.736 & 0.99 & 20 & $0.639 \pm 0.015 $& 1.473 & 1.194 & $0.763 \pm 0.018$&$-0.117 \pm 0.010$\\
R & 14.670 &  -   & -  & $1.000           $& 1.260 & 1.000 & $1.000 $&          0.000\\
I & 14.574 & 0.99 & 6  & $1.673 \pm 0.030 $& 0.902 & 0.719 & $1.203 \pm 0.022$& $0.080 \pm 0.008$\\

\hline
\end{tabular}
\end{center}
\end{table}

\begin{table}
\caption{Results for interval 8 (JD 2452800–2453100, 2003))
}
{\footnotesize
\begin{center}
\begin{tabular}{cccccccccc}
\hline
Band&$\log \nu$&$r_{iR} (1)$&n&$(F_i/F_R)^{obs}_{var} \pm 1\sigma (1)$&
    $A_{i},m$&$C_{iR} (1)$&$(F_i/F_R)^{corr}_{var} \pm 1\sigma (1)$&
    $(F_i/F_R)^{corr}_{var} \pm 1\sigma$&$\log (F_i/F_R)^{corr}_{var} \pm 1\sigma$\\
\hline
U & 14.920 & 0.77 & 25 & $0.089 \pm 0.015$ & 2.519 & 3.188 & $0.284 \pm 0.048$& $0.284 \pm 0.048$& $-0.547 \pm 0.068$\\
B & 14.833 & 0.94 &107 & $0.276 \pm 0.009$ & 1.904 & 1.810 & $0.500 \pm 0.016$& $0.500 \pm 0.016$& $-0.301 \pm 0.014$\\
V & 14.736 & 0.93 & 89 & $0.545 \pm 0.020$ & 1.473 & 1.194 & $0.651 \pm 0.024$& $0.651 \pm 0.024$& $-0.187 \pm 0.016$\\
R & 14.670 &  -   & -  & $1.000           $& 1.260 & 1.000 &  $   1.000   $   & $1.000  $        &  $0.000$\\
I & 14.574 & 0.96 &122 & $2.223 \pm 0.025$ & 0.902 & 0.719 & $1.598 \pm 0.018$& $1.598 \pm 0.018$&  $0.204 \pm 0.004$\\
K & 14.140 & 0.85 & 29 & $24.26 \pm 0.100$ & 0.171 & 0.367 & $8.903 \pm 0.040$& - & - \\
J & 14.387 & 0.97 & 38 & $0.298 \pm 0.013$ & 0.458 & 1.302 & $0.388 \pm 0.017$& $3.454 \pm 0.057$&  $0.538 \pm 0.007$\\
H & 14.262 & 0.98 & 38 & $0.604 \pm 0.018$ & 0.275 & 1.100 & $0.664 \pm 0.020$& $5.912 \pm 0.060$&  $0.772 \pm 0.004$\\
K & 14.140 &  -   & -  & $1.000          $& 0.171 & 1.000 &   $  1.000 $     & $8.903 \pm 0.040$&  $0.950 \pm 0.020$\\

\hline
\end{tabular}
\end{center}
}
\end{table}

(1) - For the $J$ and $H$ bands, relative to the $K$ band.



\begin{table}
\caption{Spectral index of the variable component for various time intervals
}
\begin{center}
\begin{tabular}{ccccc}
\hline

Interval & $\alpha \pm \sigma_{\alpha}$ & r & $\overline{F}_R$ & $F_R^{max}$\\
\hline
1, JD 2445300-2445400 & -1.88 $\pm$ 0.26 & 0.95 & 1.78 & 3.59\\
2, JD 2445600-2445800 & -0.94 $\pm$ 0.18 & 0.96 & 1.19 & 2.34\\
3, JD 2447700-2448000 & -1.42 $\pm$ 0.23 & 0.95 & 1.33 & 2.72\\
4, JD 2448220-2448350 & -2.43 $\pm$ 0.23 & 0.97 & 2.59 & 2.82\\
5, JD 2445600-2445900 & -1.56 $\pm$ 0.13 & 0.99 & 1.43 & 7.48\\
6, JD 2451000-2451200 & -1.18 $\pm$ 0.11 & 0.98 & 3.16 & 6.85\\
7, JD 2451300-2451600 & -2.29 $\pm$ 0.59 & 0.93 & 0.69 & 1.58\\
8, JD 2452800-2453100 & -2.09 $\pm$ 0.14 & 0.99 & 0.25 & 0.46\\
\hline
\end{tabular}
\end{center}
\end{table}


\begin{thebibliography}{99}

\bibitem{raiteri2001}
C.M.Raiteri, M.Villata, H.D.Aller, et al., Astron. and Astrophys.J., {\bf 377}, 396, 2001.

\bibitem{raiteri2005}
C.M.Raiteri, M.Villata, M.A.Ibrahimov, et al., Astron. and Astrophys., {\bf 438}, 39, 2005.

\bibitem{raiteri2006}
C.M.Raiteri, M.Villata, M.Kadler, et al., Astron. and Astrophys. J., {\bf 459}, 731, 2006.

\bibitem{junkkarinen2004}
V.T.Junkkarinen, R.D.Cohen, E.A.Beaver, et al., Astrophys.J., {\bf 614}, 658, 2004.

\bibitem{hagenthorn1990}
V.A.Hagen-Thorn, S.G.Marchenko, and O.V.Mikolaichiuk,
Astrofizika, {\bf 32}, 429, 1990.

\bibitem{hagenthorn1999}
V.A.Hagen-Thorn and S.G.Marchenko, Baltic Astronomy, {\bf 8}, 575,  1999.

\bibitem{hagenthorn2006a}
V.A.Hagen-Thorn, ASP Conf. Ser., {\bf 350}, 41, 2006.

\bibitem{hagenthorn2006b}
V.A.Hagen-Thorn, V. M. Larionov, N. V. Efimova, et
al., Astron. Zh. 83, 516 (2006)
                                    
\bibitem{mead1990}
A.R.J.Mead, K.R.Ballard, P.W.J.L.Brand, et al., Astron. and Astrophys. Suppl. Ser. {\bf 83}, 183, 1990.

\bibitem{wald1940}
A.Wald, Ann. Math. Stat. {\bf 11}, 284, 1940.

\bibitem{hagenthorn1990b}
V.A.Hagen-Thorn, S.G.Marchenko, L.O.Takalo et al., Astron. and      Astrophys. Suppl. Ser. {\bf 133}, 353, 1990.
\end{thebibliography}
\end{document}